\documentclass[pra,aps,preprint,showpacs]{revtex4-1}
\usepackage{graphicx}%
\usepackage{dcolumn}%
\usepackage{bm}%
\usepackage[dvips]{color}%
\def\fig_width{8.6 cm} 

\newlength{\defbaselineskip}
\newcommand{\setlinespacing}[1]%
           {\setlength{\baselineskip}{#1 \defbaselineskip}}

\begin{document}

\title{Combined ion and atom trap for low temperature ion-atom physics} 
\author{K.\ Ravi,$^1$ Seunghyun Lee,$^1$ Arijit Sharma,$^1$ G.\ Werth,$^2$ and S.\ A.\ Rangwala}
\email{sarangwala@rri.res.in}
\affiliation{$^1${Raman Research Institute, Sadashivanagar, Bangalore 560080, India}\\
$^2${Institut f\"{u}r Physik, Johannes-Gutenberg-Universit\"{a}t, D-55099 Mainz, Germany}}
\date{\today}


\begin{abstract}

We report an experimental apparatus and technique which simultaneously traps ions and cold atoms with spatial overlap. Such an apparatus is motivated by the study of ion-atom processes at temperatures ranging from hot to ultra-cold. This area is a largely unexplored domain of physics with cold trapped atoms. In this article we discuss the general design considerations for combining these two traps and present our experimental setup. The ion trap and atom traps are characterized independently of each other. The simultaneous operation of both is then described and experimental signatures of the effect of the ions and cold-atoms on each other are presented. In conclusion the use of such an instrument for several problems in physics and chemistry is briefly discussed.

\end{abstract}


\maketitle

\section{Introduction}

The cooling and trapping of ions and atoms have a number of similarities in terms of the experimental techniques and the physics they address. Individually, cold atom physics as well as cold ion physics are both sufficiently mature for investigating some of the most exciting problems, ranging from sensitive tests of QED, fundamental interactions and symmetries at one end to many particle physics at the other~\cite{Lei03,Bla10,Ye08,Leg01}. Inter-particle collisions at these temperatures exhibit several interesting features that are masked at higher temperatures/energies and is therefore likely to lead to phenomena that relate to the asymptotic molecular potential energy states. Cold and ultra-cold chemistry with molecules and atoms as the principal material ingredients is an exciting and emerging area of research~\cite{Bel09}.  

Several techniques and processes have been invented to study the low energy interactions between the constituents of dilute ultracold gases. The overwhelming majority of these employ precise addressing of the specific atomic or molecular states in order to facilitate the interactions. In a cold dilute gas of atoms, the near neighbour distances range from a few $\mu$m to a few 100 nm. These distances are very large compared to the equilibrium bond lengths of molecules, i.e. the distance that characterizes the minimum of the molecular potential energy curves/surfaces. Further, at ultracold temperatures, the atoms have negligible kinetic energy, making the collision probability for distant atoms small. Thus an important step in enhancing the reactivity between atoms at large distances is to gather them together in small local groups within the overall volume of the cold dilute gas. 

Bringing together atoms within the cold gas cloud requires an aggregating mechanism for the atoms. A natural choice for microscopic, many centered, aggregators is cold atomic ions interspersed within a cloud of dilute gas atoms or molecules~\cite{Cot02}. The Stark interaction between the ionic charge centers and the neighboring atoms has the potential to mediate the aggregation. To make this possible, cold ions must be trapped simultaneously and with spatial overlap with dilute gas atoms and molecules. Thus an important technical goal in this direction is the construction of an apparatus which is capable of holding cold ions and cold atoms simultaneously. This has been originally proposed and developed by Smith et.al.~\cite{Smi05} and later modified for specific goals~\cite{Cet07,Gri09,Zip10a,Zip10b,Sch10}.

The present apparatus is therefore designed and built with the intention of provoking chemistry of cold dilute gases, with and in the presence of ions. The ions are created from the trapped alkali atoms  in a magneto optical trap (MOT) and are optically inactive at reasonable light wavelengths and so the ions cannot be laser cooled. Several thousand daughter ions can be created and trapped from the parent MOT atoms and held in contact with the ultracold atoms. Due to the parent-daughter atom-ion relation, the resonant charge exchange channel for collision opens up, in addition to the elastic collision channel. The resonant charge exchange is important as the collision rate in this channel is independent of the collision speed of the ion atom pair over a large range of energies~\cite{Cot00}. Also the polarizability of the neutral alkali atoms in the excited states~\cite{Zhu04} is significantly larger than the ground state polarizability, adding an interesting aspect to the ion atom interaction
in a MOT. The incorporation of charged particle detection for the trapped ion(s) will allow us to detect the products of inelastic collisions and the formation of ion complexes in a generic fashion. This feature enables the measurement of branching ratios for a variety of ion-atom processes. Since the simultaneous magneto-optical trapping of various species of alkali atoms in the same experiment is relatively straightforward, cross species interactions are an important goal on the instrument described below. At temperatures lower than the MOT temperature, combining different species and isolating specific interactions in the evaporative cooling regime for atoms is expected to be very challenging as the atomic population is in constant decay and inter-atom collisions are important by themselves.  Finally, the choice of ion and atom determine the molecular states that mediate the ion-atom interaction. The alkali atom - alkali ion interact via the doublet states, the alkali atom - alkaline earth ion systems via the singlet or triplet molecular states, and so on. In each case, different physical consequences are likely.

In this paper we present our experiment and characterize the principle features of the apparatus. We discuss, (a)the design and construction of the experiment, (b) the choice of the ion and atom, (c) independent characterization of the atom and ion traps, (d) the effect of the trapping field of the ion/atom traps on the trapped atoms/ions respectively, (e) simultaneous trapping of atoms and ions, (f) the signatures of ion-atom interactions. We conclude the article with a discussion on the nature of experiments possible in the future with the apparatus described here.

\section{The Design and Construction of the Apparatus}

To accommodate trapped ions and cold trapped atoms, with spatial overlap, several aspects need to be considered. Most importantly, the effect of operation of the ion trap should have negligible effect on the atom trap and vice versa. Paul traps have no gradient magnetic field while the MOT for atoms requires the presence of a gradient magnetic field ($\approx12$ Gauss/cm). Given  a six-beam MOT along the cartesian axes, the basic design challenge reduces to construction of the ion trap such as to allow a robust MOT within its confines. Since modest electric fields are required to trap ions, there is reasonable flexibility in the design of the ion trap.

\begin{figure}
\includegraphics[width=8.3 cm]{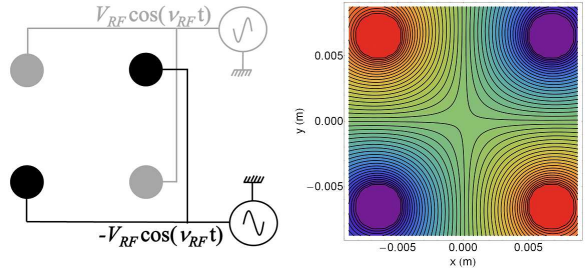}
\caption{(color online) The cross section of the ion trap is shown along with the biasing diagram and the resulting instantaneous potential.}
\label{Fig:IonTrapSchematic}
\end{figure}

The ion trap implemented in our experiment is the linear Paul trap. It consists of four parallel rods arranged in a square (quadrupole) configuration and end cap ring electrodes along the common cylindrical axis. A time varying radio-frequency (RF) voltage is applied to diagonally opposite rods, with the voltage along the two diagonals being 180$^\circ$ out of phase. The transverse cross section of the ion trap, the RF bias diagram and the potential surface at a given instant is shown in Fig.~\ref{Fig:IonTrapSchematic}. The end cap ring electrodes are held at a constant voltage with respect to the experimental ground. For such a potential an approximate analytical form for the field configuration in the central region of the ion trap can be written as,

$$ U(x,y,z,t)\approx \frac{V_{RF}}{2 r^2_0}(x^2-y^2)cos(\nu_{RF} t)+\frac{\kappa V_{ec}}{z^2_0}[z^2-\frac{1}{2}(x^2+y^2)].$$	

\begin{figure}
\includegraphics[width=7.5 cm]{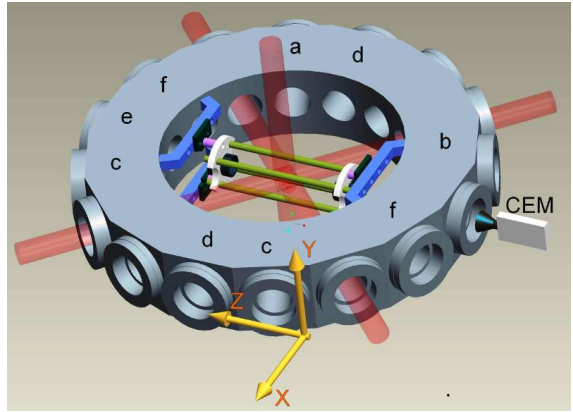}
\caption{(color online) Schematic diagram of the hybrid, ion-atom trap, within the experimental chamber. The diagram is not to scale. The ion trap axis is oriented along the z-axis. The magnetic field coils for the MOT (not shown) are external to the vacuum system, mounted symmetrically from the origin coaxial with the y-axis. A CEM for ion detection, is mounted coaxial with respect to the ion trap axis, in the -z direction from the origin. The horizontal MOT beams intersect the z-axis, at the origin, with 45 and 135 degree angles. The vertical beam intersects this arrangement orthogonally. The ports are labelled to illustrate the relative positions of the various components of the experiment as follows. (a) alkali atom dispensers, (b) the blue light source, (c) MOT imaging and fluorescence measurement, (d) Nd:YAG ionization ports, (e) the vacuum pump connection, and (f) feedthroughs.}
\label{Fig:ExptSchematic}
\end{figure}

Here $V_{RF}$ and $\nu_{RF}$ represents the applied amplitude and frequency of the time varying trapping field, $V_{ec}$, the end cap d.c. voltage, $r_0, z_0$ the trap extent and $\kappa$ is a geometric factor. For our experiment, large spacing between electrodes is required to accommodate the laser beams of 10 mm diameter to create the $^{85}$Rb MOT at the trap center, which is also the geometric center of the vacuum chamber, henceforth identified as the origin. The rod diameter `a' of 3 mm and separation of the neighbouring quadrupole electrodes `d' of 10 mm allows the 10 mm MOT beams. The end caps of the ion trap are hollow rings 25.5 mm away from the trap center. It should be noted that the RF node would intersect the MOT along a diameter of the spherical atom cloud. The schematic of the experimental arangement is illustrated in Fig.~\ref{Fig:ExptSchematic}.

Due to the unique demands of the experiment, the design of the ion trap had to be validated by numerical simulations. The potentials are generated using SIMION for each equipotential set of electrodes. Care was taken to simulate the entire potential environment in the vacuum chamber including the channel electron multiplier (CEM) so that the boundary conditions for the operating trap were realistic. 
The potential array was then imported into Mathematica, where the ion trajectories for single ions~\cite{Rav10} as well as simultaneous multiple ions can be solved for, with the desired accuracy. A fit of the above analytical potential to the numerically calculated potential yields $r_0=5$ mm. With $z_0\equiv25.5$ mm the geometric parameter $\kappa\approx 0.11$ is calculated for the trap. The extraction of the trapped ions and their detection is also simulated with accuracy.

In order to work around the above geometric constraints, and restrict to a single, flexible vacuum chamber, a Kimball Physics vacuum chamber with sixteen CF16 and two CF100 ports is used. Since the diameter of the MOT laser beams (10 mm), is small for a vapor loading MOT, the chamber was chosen to be flat. A schematic of the chamber identifying the principle components is represented in Fig.~\ref{Fig:ExptSchematic}. This particular chamber allows for large optical access in the vertical (y direction) and the horizontal (x-z plane). The ion-trap structure is mounted symmetrically, along the chamber diameter, using mounting grooves built into the chamber. This feature fixes the core of the experiment onto the chamber directly, providing rigid mounting. Electrical isolation is achieved between the various electrodes of the ion trap using a mount machined out of Macor, which is secured to the grooves in the vacuum system using groove grabbers. The metallic ion trap parts are primarily machined from SS-316 which is non magnetic and minor metallic components like screws are also non magnetic.  Electrical connections are made using multi-pin feedthroughs at the closest convenient ports. A CEM is placed, 86 mm away from the origin, along the trap axis to detect the trapped ions. Uncoated, CF view ports are used on all optically transparent ports. The MOT beam arrangement is illustrated in Fig.~\ref{Fig:ExptSchematic}. Additional view ports are used for the fluorescence detection of the trapped atoms, two-photon ionization (TPI) of the atoms by a frequency doubled, pulsed Nd:YAG laser, the mounting of getter sources for the alkali atoms, blue light LED for Light Induced Atomic Desorption (LIAD) loading of the MOT, and connecting the experiment chamber to the titanium sublimation and ion pumps. The ultimate vacuum possible in the experiment is under range of the ion gauge, and under normal operating condition the pressure is typically adjusted between few times 10$^{-10}$ mbar to 1$\times$10$^{-9}$ mbar. 

\subsection{Atom Cooling and Trapping}

The experimental arrangement for the cold atoms is typical of many vapor cell MOT apparatus and so will be dealt with very briefly. Below, we shall mostly dwell on the distinct features of the present experiment, with respect to typical MOT experiments. Here we have laser cooled and trapped $^{85}$Rb atoms, though the experiment lends itself to any other species that can be laser cooled in a vapor cell arrangement. Since the cold atoms need to be trapped at the origin, a six beam MOT is implemented, to balance the light pressure. Enough manoeuvre is available with the MOT coil mountings to enable small changes in the position of the MOT by coil adjustments. The coils comprise of 100 turns of 1.3 mm diameter wire in a 10$\times$10 cross sectional stack, wound on a copper former and is capable of sustaining over 10 Ampere continuous current without water cooling. A circulating current of 2.8 Ampere in anti-Helmoltz configuration produces a gradient magnetic(\textbf{B}) field of 12.3 Gauss/cm. Large Helmholtz coils sets along the orthogonal coordinate axes can be used to shift the magnetic center or null residual \textbf{B} fields.

The alkali atom reservoir is a Rb getter (SAES: Rb/NF/4.8/17FT10+10), which emits the atoms on resistive heating by a few Amperes of current. The MOT itself is loaded from the Rb vapor, which can either be directly emitted by the Rb getter, or alternatively, using desorbed atoms from the walls of the vacuum chamber. For the majority of the experiments reported here, we use the desorbed atomic vapor to load the MOT. To facilitate this, the chamber is illuminated by a blue light LED (Thorlabs: MRMLED) from a side port, as indicated in Fig.~\ref{Fig:ExptSchematic}. The LED illuminates over a wide angle at a central wavelength $\lambda_{LED}=456.5$ nm and $\Delta\lambda_{LED}=22$ nm, full width at half maximum. In the presence of the blue LED illumination, alkali atoms are desorbed from the cell walls due to LIAD~\cite{Meu94}, which is known to increase in efficiency with decreasing wavelength over the visible range~\cite{Ale02}. The energy of the blue light is significantly below the ionization threshold for alkali atoms, so contributes to desorption but not ionization. The desorbed atoms are much colder than the atoms emitted from the resistively heated alkali getter. Unlike the getter, the desorption method recycles the atoms already in the chamber, resulting in better vacuum. 

\begin{figure}
\includegraphics[width=7.5 cm]{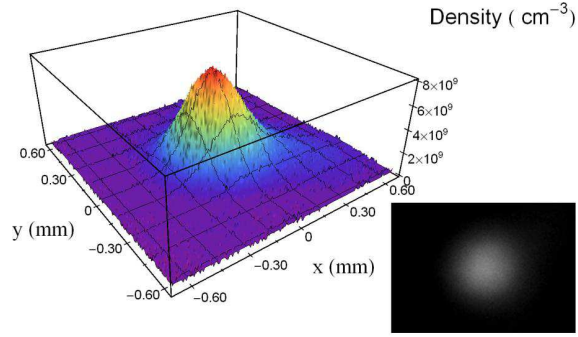}
\caption{(color online) The 2D density distribution of a section of the MOT cloud as derived from the in situ image (inset) of the fluorescence from the MOT with appropriate magnification.}
\label{Fig:MOTimageAndDist}
\end{figure}

The cooling laser light is derived from a home built extended cavity diode laser (ECDL) system capable of top of line stabilization and the repumping laser is a commercial DL100 system (Toptica). The cooling laser light is amplified using a tapered amplifier system (BoosTA) and 2.8 - 4.0 mW/beam of appropriately circularly polarized  light beams of 10 mm diameter intersect at the origin. Typically the cooling laser is detuned between 10 - 20 MHz below the $F=3\leftrightarrow{F^\prime=4}$ transition of the $^{85}$Rb D2 transition. The number of atoms that comprise the MOT varies from the high $10^5$ to low $10^6$ range. The density distribution of the MOT atoms obtained has a full width  at half maximum of $\approx{500}$ $\mu$m. An image of the $^{85}$Rb MOT, taken by a CCD camera, is shown in Fig.~\ref{Fig:MOTimageAndDist}.

The number of atoms in the MOT and their density are critical to future experiments and so is carefully measured. Fluorescence from the MOT is tracked by several CCD cameras for the position in independent directions and a spatial filtering setup in the direction shown in Fig.~\ref{Fig:ExptSchematic}, at the end of which is a photo-multiplier (Hamamatsu Photonics: R636-10). Every individual component and the entire system is carefully characterized for its transmission losses so as to determine accurately the total light flux from the MOT in the detector acceptance angle. The blue light LED is completely filtered out. Given the scattering rate of resonant light by the atoms, the atom number can be determined. The loading and decay times of the MOT are reliably measured in fluorescence, and are of the order of 10 to 20 seconds. The imaging of the fluorescence from the MOT on the CCD camera can be used to derive the density distribution of the atoms in the MOT, as shown in Fig.~\ref{Fig:MOTimageAndDist}. Careful calibration and optimization is needed to ensure that reliable numbers are measured. The principle sources of error and uncertainty for the number measurement are (a) laser intensity variations, (b) the laser detuning fluctuation and line width (c) the statistical uncertainty in the measurement of the light intensity incident on the photo-multiplier. The overall uncertainty in determining the absolute atom number using the procedures adopted is estimated to be $\pm$18\%. This measurement will be refined with time.

\subsection{Ion Trap}

The ion trap design described above is adapted to the experimental goals by enabling large optical access. In what follows we shall describe the formation, trapping and the detection of the ions in the present trap.
\\
\textbf{Formation of Ions:} The ionization energy (IE) required to just eject one electron from a Rb atom is 4.177 eV. The $^{85}$Rb$^+$ is loaded into the trap using two-photon ionization, using two different methods.\\ 
\textbf{Method (a)} Non resonant TPI. A frequency doubled Nd:YAG at 532 nm, pulse width $\approx 10$ ns, and pulse energy of the order of 165 mJ/pulse, is focused onto the trap center by a lens of focal length 150 mm, mounted outside the vacuum system. The pulse energy is adjusted such that the ionization occurs only in the focal region of the beam. The ionization can be effected from the residual vapor in the trap volume that overlaps with the 532 nm pulse, or from the atoms in the MOT. The dominant ionization process is 
$$Rb + {2^{(+)}\times{h\nu_{532}}}{\longrightarrow}Rb^+ + e^- + K.E.$$
where the superscript on 2 indicates the possiblilty of more than two photon participating in the ionization process with some probability. A single photon at 532 nm carries 2.33 eV energy, so two photons at 532 nm are sufficient to ionize a Rb atom~\cite{Tak04}. This method permits ion creation directly from background vapor, at a specific time in the experimental cycle. However, in order to load the trap from vapor, it is required to keep the getter ON, resulting in a deterioration of the vacuum in the chamber. A significant shot to shot variation, of the number of ions loaded into the ion trap, is seen in this case, irrespective of background gas or MOT loading of the ion trap.\\
\textbf{Method (b)} Resonant TPI. The two photons for ionization here come from the MOT laser (1.59 eV) and the blue light source (BLS) LED (peak at 2.72 eV) used for atomic desorption of the atoms for loading the MOT. Since the MOT lasers are either on resonance or only slightly detuned with respect to the atomic transition, this is resonant TPI. The sum of the energy carried by the red and the blue light is 4.31 eV, which is slightly higher than the IE for the Rb atom. Both the sources are continuous. In the absence of the MOT (magnetic field OFF), but MOT beams and LED ON, no ions are captured into the ion trap from the residual gas vapor, though the TPI process is active in the region of overlap of the two frequencies. In order to load the ion trap efficiently, the loading of ions must happen through the MOT atoms. The dominant ionization process is 
$$Rb + h\nu_{780}+h\nu_{456}{\longrightarrow}Rb^+ + e^- + K.E.$$ 
and thus the instantaneous velocity for the creation of the ion is almost zero. This method gives a steady loading rate of ions in the ion trap from the cooled MOT atoms. The temporal overlap of the ON phase of the ion trap along with the two light sources, loads the ion trap at a steady rate. 

Since both the above ionization schemes create ions within the trap, the electrodes of the trap can be in full trapping configuration while the ionization is underway. This is in effect a strategy which enables the loading of the ion trap at its potential minimum. The time varying RF signal produced in a function generator (Agilent: 332204) and amplified using Krohn-Hite wide band power amplifier (7602M), which generates the two 180$^\circ$ out of phase waveforms of equal amplitude, contacted to the respective diagonals of the trap. The Agilent function generator is used in the Burst Mode which is activated using a trigger pulse. This allows the RF trapping fields to be switched ON or OFF within a fraction of the RF cycle. Axial confinement is provided by a positive voltage on the end-cap electrodes. The typical range of the parameters used for trapping ions are, 400 kHz$\leq\nu_{RF}\leq600$ kHz and $0<V_{RF}\leq200$V, limited at the higher voltage and frequency values by the gain bandwidth product of the RF amplifier and $0<V_{ec}\leq75$ V. 

The prospects of fluorescence detection of the ions in the trap is compromised by the lack of convenient transitions, for the closed shell Rb$^+$ ion. However, in the event of a trapped optically active ion, for eg. a singly ionized alkaline earth atom, the optical access allowed by the experimental design permits fluorescent ion detection. For the present experiment though, only the destructive detection of the ions is implemented, using a CEM (Dr.Sjuts: KBL 10RS), which is located beyond the end-cap along the z direction as shown in Fig.~\ref{Fig:ExptSchematic}. The CEM is set up for positive ion detection and gives detectable ion signal for negative voltages greater than -1800 V at the cone of the CEM. The tail of the CEM is held at ground potential, enabling us to define V$_B$ as the bias voltage of the CEM.  The ions are extracted onto the CEM by switching the potential on the hollow ring end-cap electrode that is close to the CEM, to a negative value, while keeping the opposite end-cap electrode constant and the RF of the ion trap in steady operation. The switching of the potential on the end-cap electrode is effected within one RF cycle. The trapped ions pass through the hollow center of the end-cap electrode and are attracted by the high voltage on the CEM cone. Both the pulse counting and the analog modes of CEM operation give reliable ion signals. In the analog mode the CEM signal is fed into a low noise preamplifier (SRS: SR570), with set amplification, and is then recorded on a digital oscilloscope (DO). Due to identical charge/mass (e/M) ratios for the trapped ions, all of the trapped ions impact the detector within a time interval of 10 - 20 $\mu$s, depending on the extraction voltage. 

\begin{figure}
\includegraphics[width=7.5 cm]{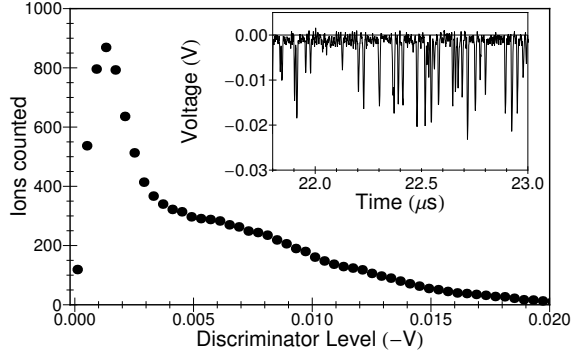}
\caption{(color online) The variation of the number of ions counted as a function of set discriminator level (DL). The inset shows a small section of the CEM signal contributing to the analysis shown in the main figure. The ideal distribution would be a monotonically decreasing number as a function of increasing magnitude of the negative DL. Every negative edge crossing the DL triggers a logical valid count and adds 1 to the total number of counts. At DL $<2$ mV, the number of counts increases because of the presence of small offset and rapidly fluctuating noise. For DL $>2$ mV, the number of counts is monotonically decreasing, as expected. However the initial decrease is rapid and is susceptible to noise counts due to baseline oscillation. A voltage level of 5 mV is identified as the discriminator level for the purpose of reliable pulse counting.}
\label{Fig:CEMdiscrL}
\end{figure}

For pulse counting detection, the entire extracted ion ensemble is allowed to fall on the CEM, which yields a 8ns, negative pulse, for every ion detected (Fig.~\ref{Fig:CEMdiscrL}(inset)). The pulse train for the 10 $\mu$s interval is recorded on a DO, and the pulses are counted by post-processing the pulse train using an appropriately set discriminator level (DL). All instances, when the falling edge of the CEM pulse train crosses the negative DL, are counted as valid counts, and signal which peaks below this level is rejected as noise. With this protocol, several ions piling up on top of each other (within 8ns), would register as a single count. From a series of experiments, we determine that $\approx300\pm30$ counts $/10$ $\mu$s is the limit of ions that can be counted. This implies a maximum counting rate of $30\times10^6$ counts/s, which pushes CEM detection to its very limit. As the ion counting is done with the recorded data train, the DL can be set very systematically as illustrated in Fig.~\ref{Fig:CEMdiscrL}. The counts as a function of DL has identifiable regions, as illustrated in Fig.~\ref{Fig:CEMdiscrL}, which allows the determination of a reasonable value of discriminator level for the analysis of the ion counts at a particular CEM bias. This procedure is benchmarked against the physical counting of the peaks in several individual time traces to make the algorithm robust against over counting or under counting. A distinct advantage of this method of post analysis is that robust and consistent algorithms can be developed for pulse counting and data from several different runs can be analyzed in a uniform manner.

\begin{figure}
\includegraphics[width=7.5 cm]{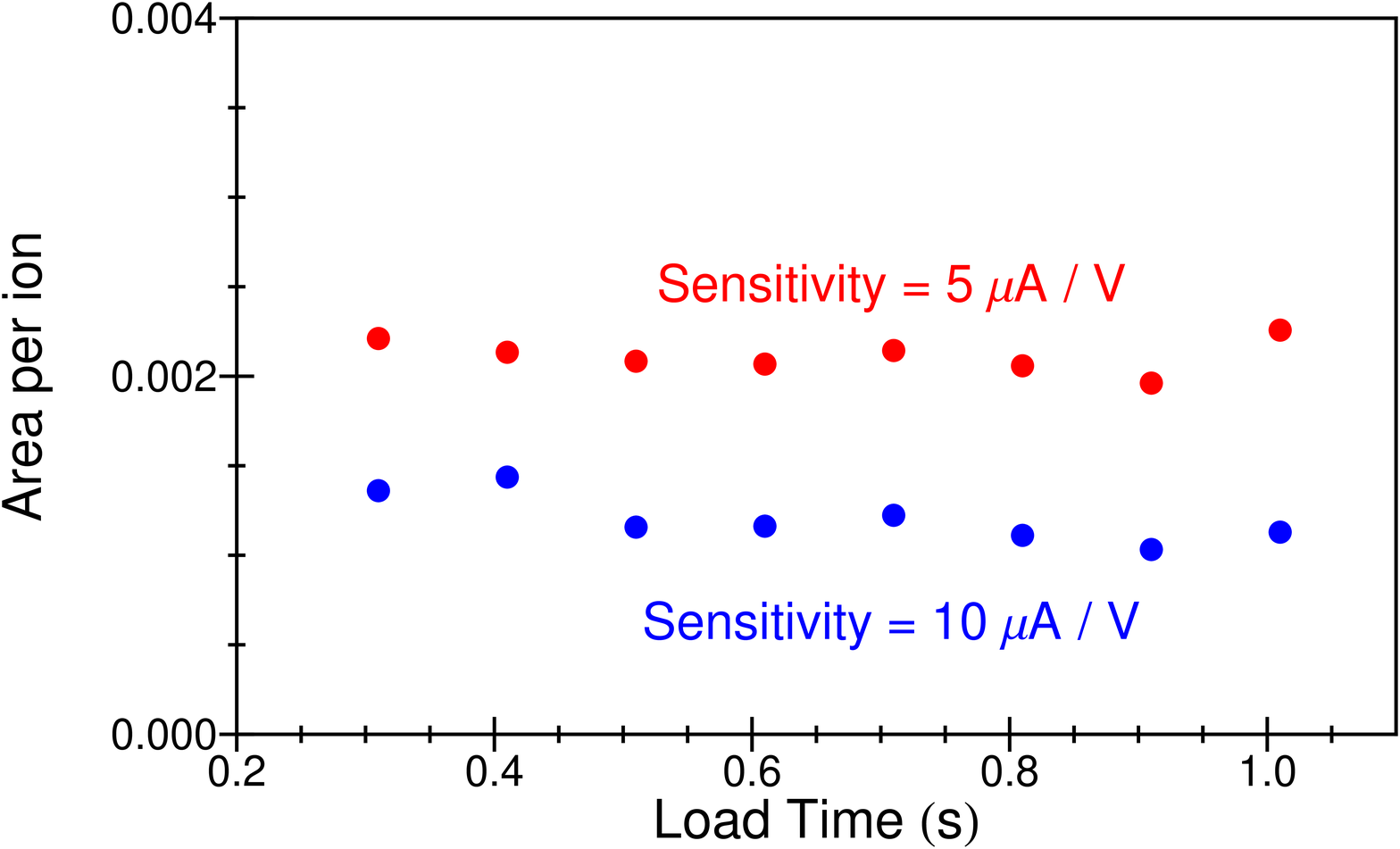}
\caption{(color online) The variation of ion counts as a function of the ion production time, using resonant TPI. The number of trapped ions produced from the MOT increases linearly with time for the first few seconds. We therefore calibrate, at $V_B=2400$ V, the correspondence between the ions counted by the pulse counting algorithm and the analog mode signal.  For 2 amplifier settings the ratio of the analog mode signal to the ion counts is plotted above. As the number of ions increases the signals are approximately constant and differ proportional to the amplifier factor. Such a calibration, allows us to switch between the two modes as per experimental convenience and extract the ion counts.}
\label{Fig:PulseVSAnalog}
\end{figure}

To reliably extend the detection ability of the CEM beyond $\approx300\pm30$ counts$/10$ $\mu$s, the integrated current mode of data acquisition described above is utilized. The ion counts in the two modes can be calibrated against the other to determine the constant of proportionality. Once this constant is measured the ion counting ability of the CEM can be reliably extended as is illustrated in Fig.~\ref{Fig:PulseVSAnalog}. Thus we have a robust characterization of the various aspects of counting ions with the CEM. Calibrating the CEM as in Fig.~\ref{Fig:PulseVSAnalog}, for different V$_B$, allows us to reliably compensate for varying detection efficiencies with V$_B$ and extend the use of the CEM for larger count rates of ions.

Having calibrated the CEM for positive ion detection, we now describe the working of the ion trap. For the experiments described below, the ions are produced by either of the two photon methods described above. Ions from hot vapor are produced by non-resonant TPI exclusively, and in most cases the ions from MOT are created using resonant TPI. Ion detection is done in either the pulse counting or the analog mode, depending on details of the experimental method. For small extraction voltages, the pulse train is spread over longer time, making it difficult to properly sample each ion count with longer time base. In such instances the analog mode is useful for capturing the data over longer time windows.

\begin{figure}
\includegraphics[width=7.5 cm]{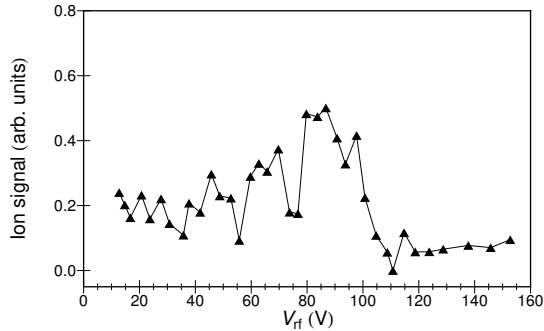}
\caption{(color online) The variation in the number of trapped ions detected as a function of the applied RF voltage at $\nu_{RF}=400$ kHz, is illustrated. The ions are created from hot vapor using non-resonant TPI. After the ionizing pulse, the ions are held in the trap for 10 ms, at a particular $V_{RF}$, and subsequently extracted to the CEM in the analog mode. Normalized ion counts are illustrated. Several  valleys are observed in the stability diagram, corresponding to ion heating resonances, produced by deviation from the ideal quadrupole potential~\cite{Dra06}. Beyond $V_{RF}=105$ V, no trapping of ions is seen.}
\label{Fig:IonsFromVapor}
\end{figure}

The trapping of the Rb$^+$ ion relies on the interplay of three parameters, the frequency $\nu_{RF}$, time varying voltage $V_{RF}$, and the static end cap voltage, V$_{ec}$. The characterization of the ion trap is achieved by changes between these three parameters. Changing the end-cap trap voltage beyond a point does not affect the number of detected ions. For most experiments reported below, the end-cap voltage, V$_{ec}=75$ V, is found to be a suitable value. Having set V$_{ec}$, we can determine the number of trapped ions ions detected vs $V_{RF}$ for different $\nu_{RF}$ values. Ionization using non-resonant TPI loads the ion trap from hot atomic vapor at the trap center. The ions were held in the trap for 10 ms and then extracted for detection. As an example the ion signal obtained is illustrated in Fig.~\ref{Fig:IonsFromVapor}, demonstrating trapping over a large variation of $V_{RF}$. The dips in the ion signal at various values of $V_{RF}$, are resonant instability points where the ion oscillation frequencies and the drive RF are linearly dependent. Qualitatively similar behaviour is seen for different values of drive frequencies.

\begin{figure}
\includegraphics[width=7.5 cm]{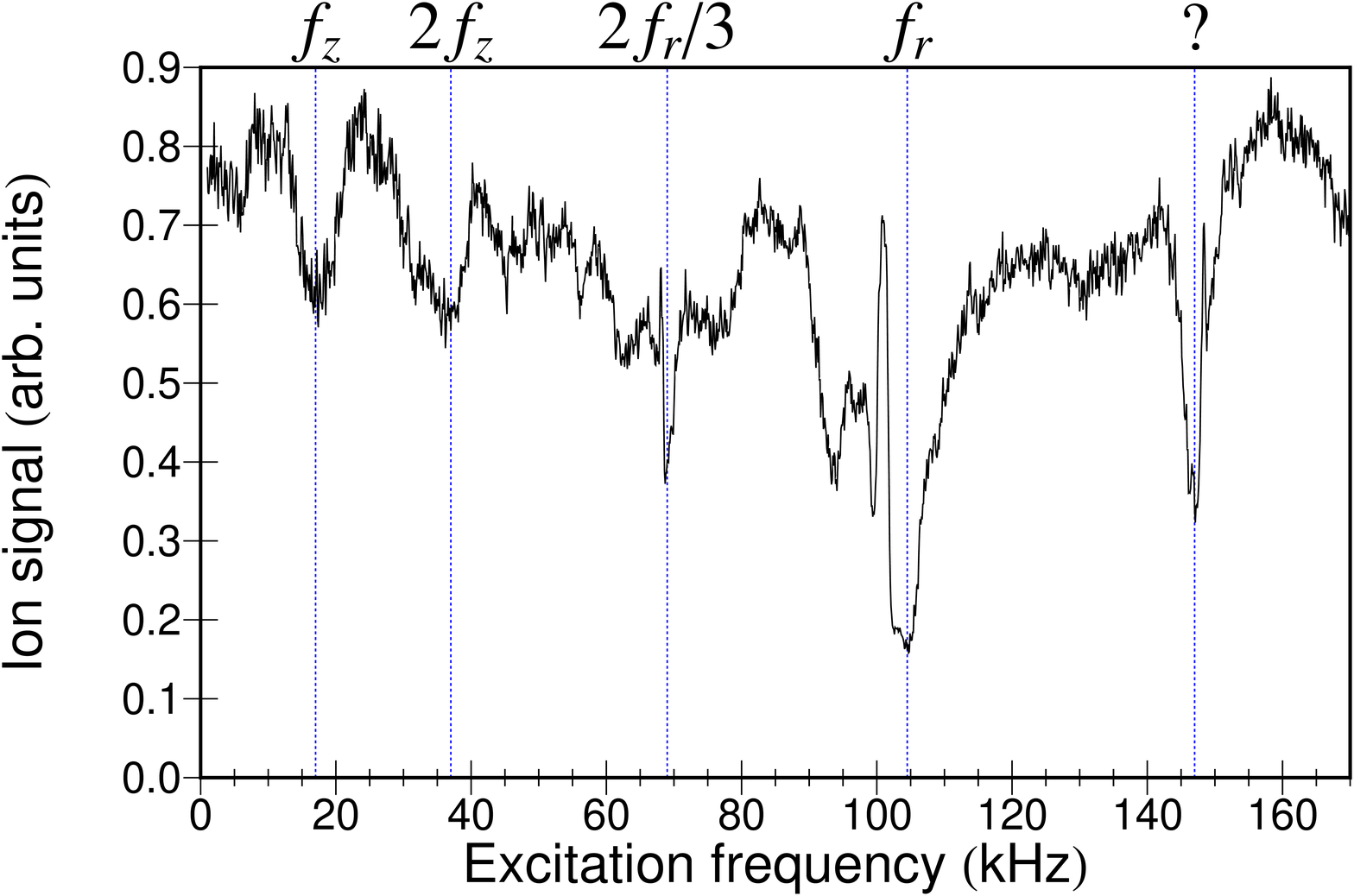}
\caption{(color online) Motional resonance spectrum of trapped Rb$^+$ ions. For this measurement, $V_{RF}=100$ V, $V_{ec}=75$ V, and $\nu_{RF}=500$ kHz. The hold time for the trapped ions is 10 ms, during which a low power excitation RF voltage, is mixed in with the $V_{RF}$ and applied to the quadrupole electrodes. The figure illustrates the variation of the ion counts as a function of the excitation frequency. Each point represents a full cycle of ion creation, trapping and excitation, and ion detection. The resonance at $f_r\approx$ 100 kHz is identified as the fundamental radial resonance. As the excitation is fed in through the quadrupole rods, the radial motion couples easily to the excitation, and so manifests at low excitation powers. The major resonances are labelled and the narrow resonance at 146 kHz is unidentified.}
\label{Fig:SecFreq}
\end{figure}

Since the ions are trapped dynamically, along with the frequency of the micromotion, which is a result of forced oscillation tied to the applied RF field, a slower, orbital macromotion manifests for the trapped ions. Experimental determination of the macromotion frequencies is done by applying a weak monochromatic perturbation, which can excite the motion of the ions in the trap, in the vicinity of the natural frequencies of the ion motion. The nature of the resonances for specific trap parameters, is illustrated in Fig.~\ref{Fig:SecFreq}, for ions from hot vapor. Several resonances are seen at the  harmonics, sub-harmonics and the natural frequencies of oscillation in x-y plane.  Since our perturbation frequency is applied radially, it is much harder to couple the excitation to the z motion, and so this motion requires more power to excite. The numerically computed values of the trap frequency agree very well with the experimental values. These are found by taking the Fourier transform of the trapped ion trajectory for the experimental trap parameters.

\begin{figure}
\includegraphics[width=7.5 cm]{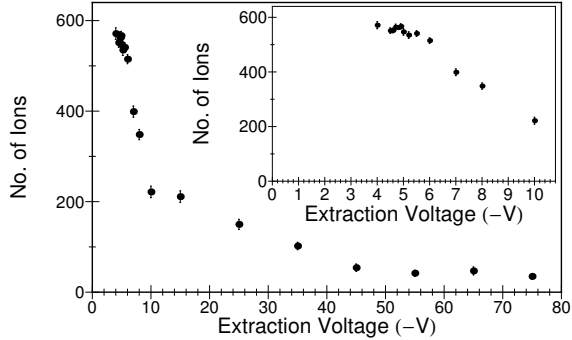}
\caption{(color online) The ion counts recorded in the CEM as a function of the end-cap electrode extraction voltage ($V_{extr}$), on the CEM side. Here $V_B=2400$ V, and the electrode at the far end is held at +75 V. Ions are loaded into the ion trap for identical duration using resonant TPI. The end cap electrode voltage on the CEM side is switched from +75 V to $-V_{extr}$ V and the ion number counted clearly increases as $\left|V_{extr}\right|$ goes to smaller values.}
\label{Fig:IonCountsWithExtrVolt}
\end{figure}

Finally it is important to understand how to lose minimum number of trapped ions in the process of extracting them onto the detector. Both simulation and experiments confirm that the number of ions detected increases with decreasing value of the switched end-cap voltage, due to lensing effects. This is illustrated in Fig.~\ref{Fig:IonCountsWithExtrVolt}. 

\section{Inter-Trap Interactions}

Vital to the smooth function of the hybrid trapping scheme is the interaction of the trapped species (ion or atom) with the empty but fully operational other trap (atom or ion). To check this experimentally, both traps are ON but only one is loaded. 

\begin{figure}
\includegraphics[width=7.5 cm]{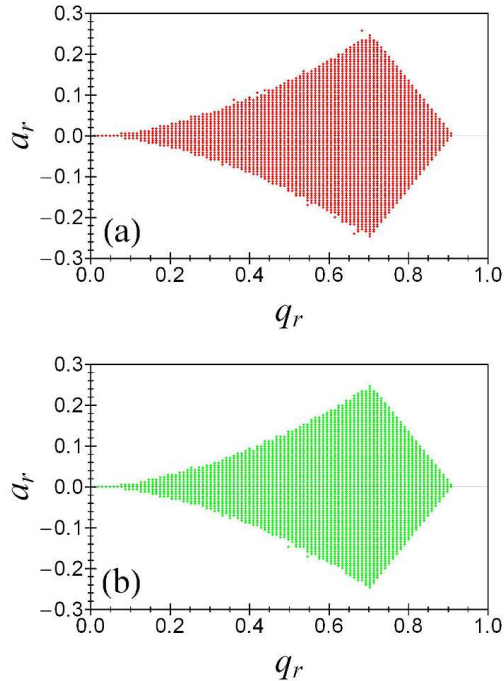}
\caption{(color online) The calculated $a_r- q_r$ stability diagram for ions in the trap potential. Filled points indicate the region of stability. (a) is the stability in the absence of the magnetic field and (b) with a gradient quadrupolar magnetic field of 24 Gauss/cm. The regions of $a_r- q_r$ space that support trapping of an ion are almost identical.}
\label{Fig:XYstabWandWoutB}
\end{figure}

The trapped ions see the laser beams at 780 nm and the inhomogeneous gradient magnetic field for the MOT. The optical frequencies are too detuned from any ion transition to have a significant effect on the ions. The inhomogeneous gradient magnetic field however changes the ion trajectories only marginally, because of the Lorentz force. Simulations show that this field makes only very slight change to the stability of the trapped ions in the $a_r - q_r$ plane, where $a_r$ and $q_r$ are scale parameters in the Mathieu equation, the solution to which determines the ion trajectory. The primary region of stability for a single ion in the $a_r - q_r$ plane is calculated for the no magnetic field case and an externally imposed quadrupole magnetic gradient field of 24 Gauss/cm (this field gradient is approximately twice the usual gradient field used for the MOT). The region of stability for the two cases is shown explicitly in Fig.~\ref{Fig:XYstabWandWoutB}. Indeed experimentally no systematic change in the number and the duration of the trapped ions is seen, thus confirming the conclusions of the simulation. To probe the effect of the magnetic field on the motion of the ion experimentally, a motional resonance experiment was performed to investigate any deviation from the spectrum in Fig.~\ref{Fig:SecFreq}. However no experimentally significant deviation was seen in this measurement when compared with the data in the absence of the MOT magnetic field in Fig.~\ref{Fig:SecFreq}. 

The principle source of perturbation for the MOT in the presence of the ion trapping fields is the RF. The typical operating frequency for the ion trap is in the range of a few 100 kHz. In this range of frequency the cold atom at the center of the trap are in principle, vulnerable to Zeeman transitions within the respective states present at the trap center. This has been experimentally investigated by tracking any change in the MOT fluorescence as a function of applied RF and end-cap voltages. No detectable change is observed. In addition the MOT lifetime is also unchanged in the presence of the RF fields indicating that there is no detectable heating of the MOT by ion trap operation. 

To summarize, while the operation of the extra trap has the potential to affect the trapped ions or atoms, in the regime of the present experiment these inter trap effects can be generally neglected, without serious consequences.

\section{Ion-Atom Trapping and Interactions}

Loading the ions from the cold atoms requires an experimental choice between the two ionization methods discussed above. It is experimentally found that the combination of cold atoms with the BLS illumination builds the detectable ions in the ion trap at a steady rate. The formation of ions from the MOT atoms is directly related to MOT density over the MOT volume. In addition, the frequency and intensity stability of light sources is uniform with time, and a multi-photon process with only one of the two light sources in operation is extremely improbable. On the other hand the Nd:YAG ionization suffers from many uncertainties. These include, shot to shot intensity fluctuation, uncertainty over the volume of ionization from the MOT, the precise order of the multi-photon process resulting in ionization, etc. Thus for all the ion atom combination experiments that follow, the ion trap is loaded from the MOT atoms using resonant TPI. 

\begin{figure}
\includegraphics[width=7.5 cm]{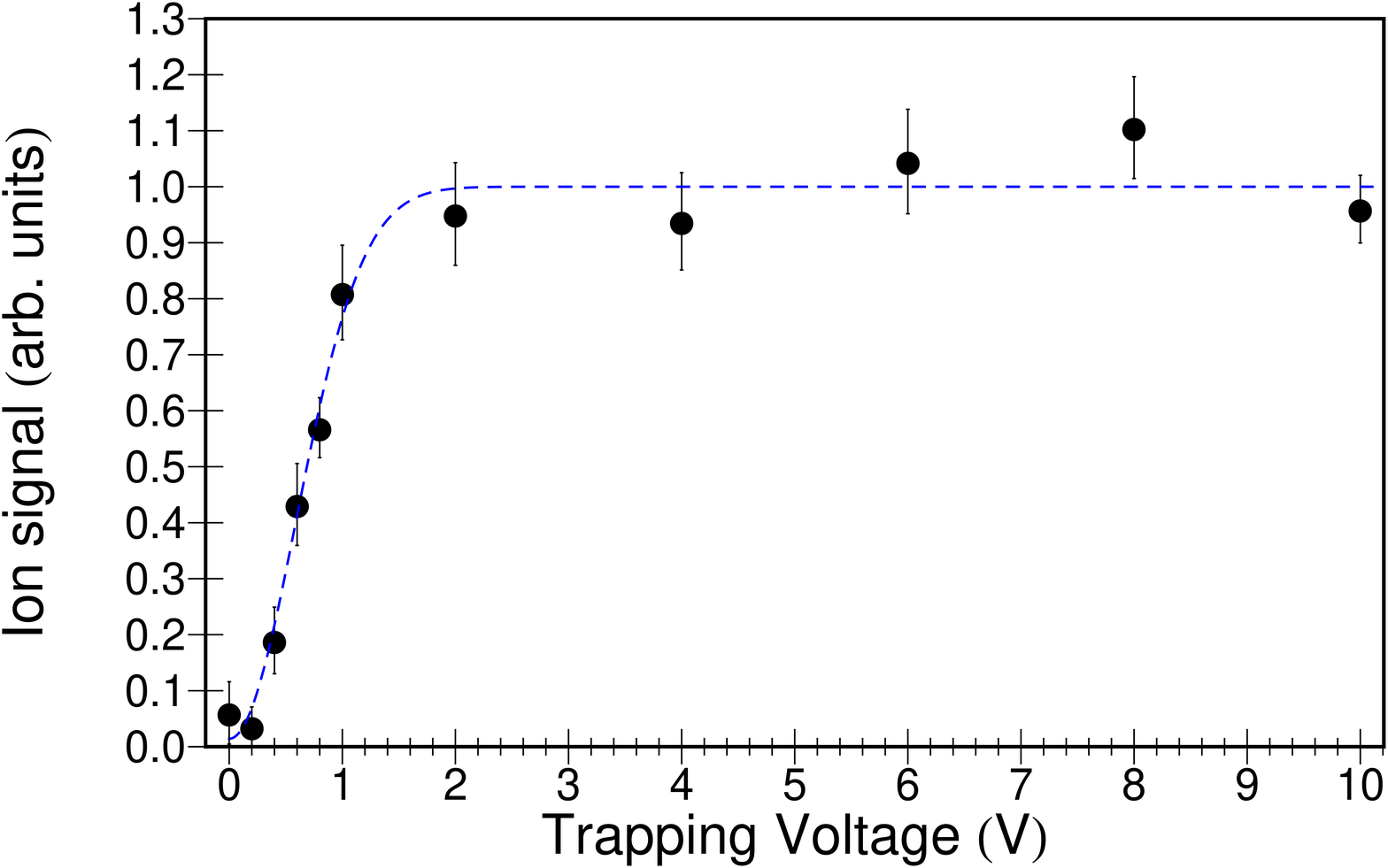}
\caption{(color online) The number of ions detected from the ion trap as a function of the trapping voltage, $V_{ec}$. The axial direction represents the direction of weakest confinement for the ions. The ion counts are normalized to emphasize a fully loaded trap and an empty trap. The loading time is kept at 0.2 s so that the ions do not have too much time to interact with the cold atoms, while generating a reasonable ion number in the trap volume. An inverted Gaussian is fitted to the data and is meant to guide the eye}
\label{Fig:IonTemp}
\end{figure}

Since the MOT temperatures are of the order of $\approx100$ $\mu$K range and the recoil imparted to the atom in the process of ionization small, the ion  created from the MOT has very little kinetic energy. However because the ions are created over the spatial extent of the MOT, they can be created with significant potential energy depending on where in the trap potential and which time in the RF cycle they were ionized. For this reason it is prudent to keep the size of the MOT small so that the ions are created in the close vicinity of the RF node. Within these constraints a temperature measurement can be made for the ions loaded into the trap as shown in Fig.~\ref{Fig:IonTemp}, which is made as follows. For the instance when $V_{rf}=96$ V and $\nu_{rf}=400$ kHz and with $V_{ec}$ set to a specific value between 0 and 10 V, the MOT is loaded to saturation, the ion trap switched ON for 0.2 seconds during which the the ion trap loads. After the 0.2 seconds of loading the end-cap voltages, are switched to their respective extraction values of $V_{ec}$ and -5 V, and the ions extracted with the cycle repeating. Several measurements are made for each end-cap trap voltage value for better statistics. Fig.~\ref{Fig:IonTemp} illustrates the variation of the ion counts with $V_{ec}$ giving an indication of the temperature of creation of ions from the MOT, in the hybrid trap. For trap depths below 1.5 eV the ions start to leak out of the trap. While the present implementation of the trap temperature measurement is along the z direction, it is sufficient to determine the ion temperature. This is because for long hold times or with many ions present, ion-atom and ion-ion interaction will ensure equilibration. In what follows we discuss some experimental signatures of precisely this ion-atom interaction.

\begin{figure}
\includegraphics[width=7.5 cm]{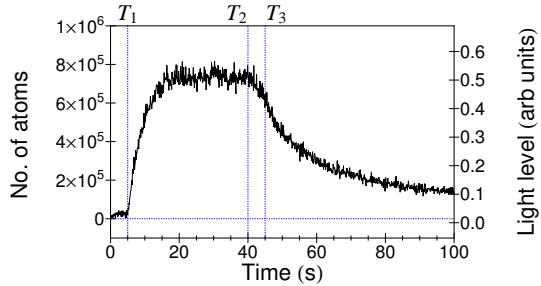}
\caption{(color online) The time sequence of a typical ion-atom experiment. The figure shows the evolution of the atom fluorescence at different stages of the experimental cycle as described in the text. The duration from $T_2$ to $T_3$ seconds is when the MOT and the ion trap are simultaneously loaded. The resulting drop in the atomic fluorescence is clearly manifested. Beyond $T_3$, the MOT and the ions in the ion trap are allowed to decay naturally. The vertical left axis indicates the total atom number loaded in the MOT, arrived at by the procedure described earlier. The vertical right axis indicates the light level detected by the PMT.}
\label{Fig:MOTpIonLandD}
\end{figure}

To investigate ion-atom interactions, a typical experiment is executed as follows. The MOT starts loading, when the blue LED is turned ON at time $T_1$, from the vapor in the vacuum system that is generated through LIAD, as shown in Fig.~\ref{Fig:MOTpIonLandD}. The ion trap end-caps are kept ON while the RF is OFF. All the while the MOT is loaded, the blue LED is creating ions from the cold atoms by resonant TPI, as described earlier. However these are not trapped and escape the trap region instantly. Once the MOT is loaded to saturation, the RF is switched ON at time $T_2$, thus turning ON the ion trap. Immediately the ion trap starts loading resulting in ions and atoms trapped in the same volume. As soon as the trapped ions coexist with the trapped atoms, the MOT fluorescence drops by $\approx20$\% as indicated in Fig.~\ref{Fig:MOTpIonLandD}. The resonant TPI rate does not change with the switching ON of the ion trap RF, as it depends only on the MOT density, MOT lasers and the blue light LED being on simultaneously. The drop in fluorescence from the MOT is a distinct signature of ion-atom interactions, between the MOT atoms and the trapped ions. Collisional interaction can be elastic and enhanced by resonant charge exchange between the MOT atoms and the trapped ions, causing the loss of cold atoms from the MOT. However, there also exists the possibility of molecule/molecular ion formation, due to the combination of ions, atoms and several different light fields being present all at once. The details of the loss mechanism will be explored in detail in future experiments. A few seconds after the RF is switched ON, the blue light LED is switched OFF ($T_3$), resulting in the general decay of the MOT and the trapped ions. 

\begin{figure}
\includegraphics[width=7.5 cm]{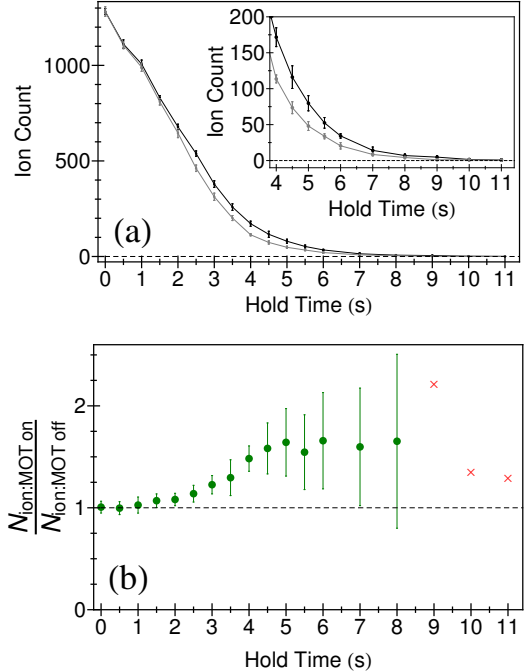}
\caption{(color online) The decay of Rb$^+$ ions from the ion trap in the presence and absence of an overlapping MOT. The MOT is loaded from $T_1$ to $T_2$, the ion trap is loaded for the time interval $T_2 < T_l < T_3$ and beyond $T_3$ the MOT is either held constant or switched OFF. (a) illustrates the decay of ions from the ion trap as a function of hold time beyond $T_3$ in the presence of a constant MOT (Black) and without the MOT (gray). The inset in (a) is a zoom in for longer hold times. From (a) it is clear that the number of ions held in the presence of the MOT is systematically higher than without the MOT. The ratio of the ion counts in the two cases is illustrated in (b), where by 4.5 s of hold time the ratio of ions with MOT to without MOT has stabilized to $\approx 1.61$. Beyond 8.0 s the error bars are not shown in (b) as the statistical scatter in the ion number held without the MOT goes below the mean value of the baseline count for the measurement and the ratio at 9 s is divided by 2 to avoid scale compression.}
\label{Fig:IonLifeCombin}
\end{figure}

Following the experimental signature of ion-atom interaction on the atom signal we demonstrate a corresponding signature in the ion signal. Here we measure the lifetime of the ions in the ion trap, with and without an overlapping MOT. For this experiment, the MOT is loaded with the dispenser ON and the BLS is turned ON for the ion trap loading time $T_l$, from $T_2$ to $T_3$. Following the loading of ions at $T_3$, the number of ions as a function of hold time is measured, either in the presence of the constant MOT or with the MOT switched OFF at $T_3$. The results of this experiment are illustrated in Fig.~\ref{Fig:IonLifeCombin}(a) and (b). Ten independent measurements of ion number are performed at each hold time and the statistical uncertainty is one standard deviation. It is found that identical numbers of ions are held in the ion trap at $T_3$. As the hold time increases, while the overall ion numbers in the trap decreases both with and without a MOT, the number of ions in the trap with simultaneous MOT is systematically higher on the average by $\approx 61\%$ as in Fig.~\ref{Fig:IonLifeCombin}(b).

\begin{figure}
\includegraphics[width=7.5 cm]{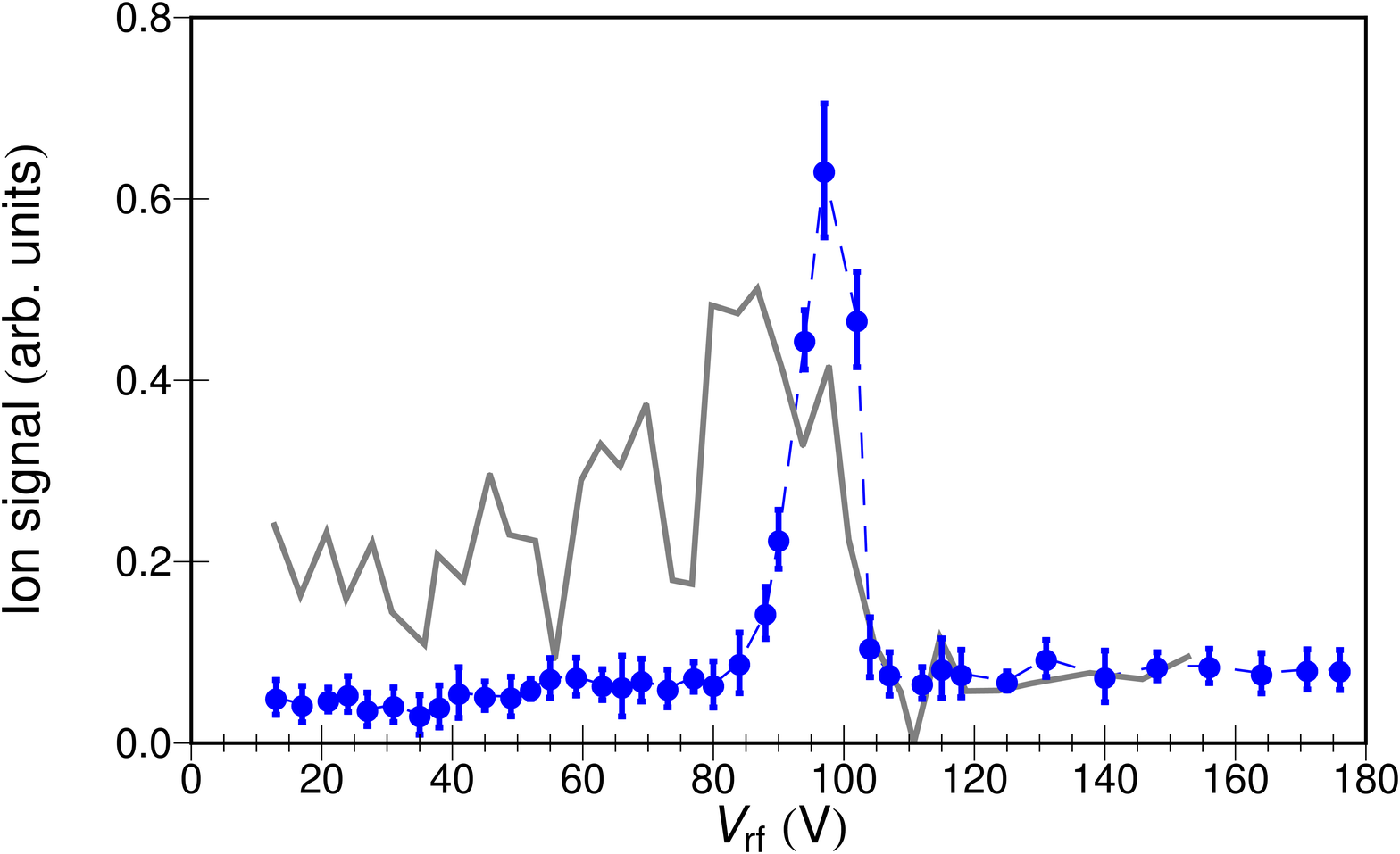}
\caption{(color online) The variation in the number of trapped ions detected, in the presence of a MOT, as a function of the applied RF voltage with $\nu_{RF}=400$ kHz is illustrated. The equivalent result with only ions is superimposed for comparison (Fig.~\ref{Fig:IonsFromVapor}). The ions are created from cold atoms using resonant TPI. After the ionization part of the cycle, the ions are loaded into the trap from the MOT for 510 ms, at a particular $V_{RF}$, and subsequently extracted to the CEM. The ion numbers are normalized.  The ion trap is found to work in a much narrower range of $V_{RF}$ in the presence of cold atoms.}
\label{Fig:StabIonWAtoms}
\end{figure}

Finally we look at the stability of the ions in the ion trap, in the presence of the MOT atoms, analogous to Fig.~\ref{Fig:IonsFromVapor}. Here, the ions are loaded from the MOT by resonant TPI for 510 ms and then extracted. The range of V$_{RF}$ values for which ions are seen is dramatically reduced as compared to when there are no cold atoms. A comparison of these two instances are shown in Fig.~\ref{Fig:StabIonWAtoms}, where only the highest values of V$_{RF}$ support ion trapping in the presence of the MOT atoms. This is in contrast with the relatively broad regions of trapping, for the same frequencies, in the absence of cold atoms as shown in Fig.~\ref{Fig:IonsFromVapor}. Whether the narrowing is due to direct ion-atom collision is to be investigated in the future. 

Thus we have demonstrated that both cold atoms and ions can be trapped simultaneously, and with spatial overlap. A few distinct signatures of the ion-atom interaction that results have been presented. This validates the instrument concept, design and the techniques employed here. Below we discuss in brief the physics problems that can be addresses with such an instrument, which shall put the instrumentation effort in perspective.  

\section{Conclusions and Science Directions}

In conclusion, we have designed, built and characterized an apparatus that allows the simultaneous trapping of cold atoms, and ions. The focus of the manuscript has been to characterize in detail the unusual aspects of the present experiment and the innovations developed for this purpose. Techniques that are standard have been dealt with only briefly, to the extent they serve to complete the picture. The range of experiments described cover the operation of individual traps, inter-trap perturbation, combined operation of the traps and the ion-atom interactions observed.

The experiment described above opens up the possibilities of several investigations of ion-atom interactions over mK $\leq T\leq\ 10^3$ K temperatures. This can be done, as the ions created from the MOT, can over time be cooled by the ultra-cold atoms. The atoms and ions can be loaded in their respective traps to steady state populations, i.e. trap saturation. Questions such as (a) determination of the ion-atom collision cross sections as a function of temperature, (b) how to sympathetically crystallize ions, (c) what is the combined temperature of the ion-atom system~\cite{Mak03}, etc, can be posed. In addition (d) what are the reaction products of of the ion-atom-photon soup, (e) is photo-association aided, (f) can macro-molecules be formed, and so on are other experiments possible. Further, more than one species of ion and/or atom can be loaded simultaneously and the cross species processes can be studied, which would open up a larger canvas for exploration. In particular, we emphasize that these experiments can be done in steady state as the ion and atom numbers can be held constant, by loading their respective traps to saturation. Our experiment therefore allows studies with large numbers of ions and atoms in a wide temperature range, with flexibility to increase the complexity in a straightforward manner.  It might well be that such a system finds its principle utility as a `test tube' for low temperature processes, that does not depend on specific state addressing. It is for the future to reveal what emerges from this `test tube'. The instrument and experiments described here enable the systematic study of many questions, including those listed above.

\section{Acknowledgments}

The authors would like to acknowledge the excellent technical support provided by the RRI machine shop and its staff for the fabrication of this experiment. Crucial electronics fabrication and support by S. Sujatha from RAL at RRI is gratefully acknowledged. Tridib Ray's involvement with the project has been of great value. Finally the authors are happy to acknowledge constructive discussion and incisive comments from Eric Cornell on many aspects of the present experiment.

\end{document}